\documentclass[structabstract]{aa} 
\usepackage{graphicx}
\usepackage{txfonts}
\begin{document}
   \title{The spectroscopic  evolution of the symbiotic star AG Draconis}
   \subtitle{I. The O VI Raman, Balmer, and helium emission line variations during the outburst of 2006-2008}

   \author{S. N. Shore\inst{1,2}, G. M. Wahlgren\inst{3,4}, K. Genovali\inst{1}, \\ 
   S. Bernabei\inst{5},  P. Koubsky\inst{6}, M. \v{S}lechta\inst{6}, P. \v{S}koda\inst{6},  A. Skopal\inst{7}, and M. Wolf\inst{8}}

  \institute{
   Dipartimento di Fisica ``Enrico Fermi'', Universit\`a di Pisa, largo B. Pontecorvo 3, I-56127 Pisa, Italy \email{shore@df.unipi.it}
   \and
   INFN - Sezione di Pisa
   \and  
   Catholic University of America, Dept. of Physics, 620 Michigan Ave NE, Washington DC, 20064, USA  
   \and
NASA-GSFC, Code 667, Greenbelt, MD, 20771, USA 
\and
INAF - Osservatorio Astronomico di Bologna, via Ranzani 1, I-40127 Bologna, Italy
 \and
  Astronomical Institute, Academy of Sciences of the Czech Republic, Ond\v{r}ejov, Czech Republic 
  \and
  Astronomical Institute, Slovak Academy of Sciences, 059 60 Tatransk«a Lomnica, Slovakia  
\and
Astronomical Institute, Faculty of Mathematics and
             Physics, Charles University Prague, CZ-180~00 Praha 8, 
             V~Hole\v{s}ovi\v{c}k\'ach 2, Czech Republic}
             
              \date{Received ---; accepted ---}
 
  \abstract
   {AG Dra is one of a small group of low metallicity S-type symbiotic binaries with K-type giants that undergoes occasional short-term outbursts of unknown origin. }
   {Our aim is to study the behavior of the white dwarf during an outburst using the optical Raman lines and other emission features in the red giant wind.  The goal is to determine changes in the envelope and the wind of the gainer in this system during a major outburst event and to study the coupling between the UV and optical during a major outburst}
   {Using medium and high resolution groundbased optical spectra and comparisons with archival $FUSE$ and $HST/STIS$ spectra, we study the evolution of the Raman O VI features and the Balmer, He I, and He II lines during the outburst from 2006 Sept. through 2007 May and include more recent observations (2009) to study the subsequent evolution of the source.}
   {The O VI Raman features disappeared completely at the peak of the major outburst and the subsequent variation differs substantially from that reported during the previous decade.  The He I and He II lines, and the Balmer lines, vary in phase with the Raman features but there is a double-valuedness to the He I 6678, 7065 relative to the O VI Raman 6825\AA\  variations in the period between 2006-2008 that has not been previously reported.  }
   {The variations in the Raman feature ratio through the outburst interval are consistent with the disappearance of the O VI FUV resonance wind lines from the white dwarf and of the surrounding O$^{+5}$ ionized region within the red giant wind provoked by the expansion and cooling of the white dwarf photosphere.}


   \maketitle

\section{Introduction}


Although as a class, the symbiotic stars are enigmatic, the S-type star  
AG Draconis is unusually full of surprises.  The system has an orbital period of approximately 550 days (Meinunger 1979, Fekel et al. 2000) and consists of a low metallicity K0-K3 giant (see, e.g., Smith et al. 1996) 
and a hot white dwarf (WD), whose effective temperature is 
estimated to be between 100kK and 170kK.   Although there are no eclipses, the ultraviolet (UV) variations are consistent with orbital modulation due to wind absorption of the WD continuum and its associated ionized region by the giant  (Gonzalez-Riesta et al. 1999, hereafter G99; Young et al. 2005).   The archival photometric history of AG Dra is particularly interesting.  For instance, ten major outbursts have been recorded for AG Dra, with a $V$ \ of 2 mags, between 1890 and 1960 (Robinson 1969) at intervals of roughly 14 to 15 years (Viotti et al. 1983; Mikolajewska et al. 1995;  Tomov \& Tomova 2002; Leedj\"arv et al. 2004, hereafter L04; G\'alis, Hric, \& Petr'k, K. 2004; Viotti et al. 2007, Skopal et al. 2007). The major events are double-peaked, with interburst intervals of $\approx$1 yr.  Since the outburst of 1980, additional bursts of weaker magnitude have been recorded that show a similar inter-outburst timescale.  It is the strongest ``supersoft'' X-ray source known among the symbiotics (Greiner et al. 1997, Gonzalez-Riestra et al. 2008, hereafter G08) and is strongly variable with two characteristic states of hardness.  Gonzalez-Riestra et al. (1999) distinguish between ``cool'' and ``hot'' outbursts, where the major outbursts of 1980 and 1994 were ``cool''  according to the He II Zanstra temperatures.  These were  interpreted as a cooling of the WD as its radius increases, leading to higher luminosity with envelope nuclear burning.  Observations of AG Dra taken over the lifetime of $IUE$, and including the outburst episodes of 1980 and 1994, have been published by Gonzalez-Riesta et al. (1999). They note that the long wavelength UV continuum flux decreased by 20\% over the observation baseline.  Zanstra temperatures are in the range 90kK to 110kK.  Eriksson et al. (2006) determined WD wind velocities of 120 to 150 km s$^{-1}$ based on a re-analysis of the $IUE$ high resolution spectra.   

There is also a significant far ultraviolet flux (FUV) archive for AG Dra ($\lambda < 1200$\AA\ ), virtually all of it from quiescent intervals.  $ORFEUS$ spectra, obtained in 1993 with $BEFS$, show a broad O VI 1031 \AA\ 
line with peak flux of 3$\times$10$^{-12}$ erg s$^{-1}$cm$^{-2}$\AA$^{-1}$ 
along with strong narrow red giant wind components ($>$10$^{-11}$). $TUES$  
spectra from 1996 showed a broad O VI line with peak flux 4$\times$10$^{-12}$ 
(Schmid et al. 1996). $HUT$ spectra obtained in 1995 show the broad line with peak flux 6$\times$10$^{-12}$ but at comparatively low resolution (Birriel, Espey, \&  Schulte-Ladbeck 2000).  
$FUSE$ spectra were taken during quiescent periods  from March 2000
to December 2004 and in April 2007 during the decay of the 2006 outburst.
In these data, Young et al. (2006) identified emission  lines from highly ionized species, 
consistent with an electron temperature $T_e \approx 20 - 30$ kK. 
The high ionization Ne lines indicate a higher temperature closer to the WD. 
The broad O VI line was clearly variable even during periods with the
same $U$ magnitude; the difference appears to be when the observations
occurred relative to the outbursts and the type of outburst. 
Spectra were obtained with a range from ($<$0.5 to 5)$\times$10$^{-12}$, 
the weak line occurred at the last epoch. Notably, $BEFS$ and $FUSE$ detected 
P Cyg components from the K star wind whenever the broad line was
strong. One $HST/STIS$ Ly$\alpha$ spectrum shows a broad, strong emission
profile with P Cyg absorption at greater than $-2000$ km s$^{-1}$ at about the
same time as the 2003 $FUSE$ spectrum, a feature unobservable with $IUE$.
The Lyman decrement is very large: no emission is ever detectable at
Ly$\gamma$ and it is weak at Ly$\beta$ in all FUV spectra.

The latest major outburst of AG Dra began in late August 2006 (around MJD 53900), 
earlier than anticipated from the average time between major outbursts, and
reached a maximum of $V \approx$ 8. This outburst was followed 
by a second, smaller one in October 2007 that showed a more rapid decline and 
reached a maximum $V \approx$8.8. 
Unlike the previous two major outburst events, the UV spectrum 
(900 to 3000 \AA) was not accessible with spectroscopy.   However, the optical O VI $\lambda\lambda$6825, 7080\AA\ features are formed by Raman scattering on Ly$\beta$ of the FUV O VI resonance doublet.  As such, they provide a window into the behavior of the WD during an outburst that is not affected by orbitally modulated absorption.   AG Draconis is particularly notable for the extreme conversion efficiencies previously derived for the Raman scattering, of order 50\% (Schmid et al. 1999; see, however, Birriel et al. 2000).  With our optical spectroscopic monitoring, which began at the peak of the 2006 outburst and 
continues to the present, we can correlate the behavior of spectral variations 
and quantify flux variations in the UV from the peak of the outburst through its subsequent decay.  Here we concentrate on the most significant new result,  the variation of the O VI Raman features during the outburst, how it compared with the helium and Balmer line variations, and what it may reveal about the high energy continuum and activity of the WD.


\section{Observations}

Our optical observational data set consists of spectra taken with the Cassini 1.5m at Loiano Observatory with all four grisms (covering 3700 - 8500\AA, resolution between 0.4\AA to 4\AA\ per pixel, exposure times from 10 to 1200 sec), and the 2.0m Ond\v{r}ejov Observatory with the RETICON spectrograph (6400 - 6950\AA, 
resolution 0. 24\AA\ per pixel, exposure time was typically 1200 sec).  The journal of observations and the tables of measurements are included in the online material.  All spectra were reduced using IRAF and our own special purpose routines written in IDL.\footnote{ IRAF is distributed by the National Optical Astronomy Observatories, which are operated by the Association of Universities for Research in Astronomy, Inc., under
cooperative agreement with the US National Science Foundation.}  In several instances, contemporaneous spectra allow us to correlate and cross-calibrate the data from the three sites.   The continuum variations were checked based on measurements of absorption line equivalent widths from the K star photosphere and the fluxes were corrected for continuum variations using contemporary CCD photometry.   

For phasing we have used the orbital ephemeris of Fekel et al. (2000) based on the K star radial velocities,
\begin{displaymath}
T(v_{\rm rad, max})  = 2450912.5 (\pm 4.1) +548.^{d}65 (\pm0.97)E
\end{displaymath}
with an amplitude of 5.86$\pm$0.30 km s$^{-1}$.  This velocity amplitude is resolved only in our TNG spectrum and confirmed by cross correlation of the Ond\v{r}ejov data as an upper limit ($\le 1.7$ km s$^{-1}$ using cross correlations at 6540-6550\AA).

These new observations were supplemented by archival spectra.   An archival $HST/STIS$ medium resolution observation covering much of the UV and optical (1150 - 8000\AA\ with some gaps in spectral coverage) is available (program O6YK, 2003 Apr. 19, MJD 52748) at orbital phase 0$^p$.34.   Wavelengths were checked for the $HST$/STIS spectra and registered using the interstellar velocities measured for the Na I D lines and the published $IUE$ velocities.  The $FUSE$ spectra were wavelength corrected using the interstellar H$_2$ absorption lines.    We also used a Telescopio Nazionale Galileo (TNG) spectrum, obtained at high resolution on 2005 Aug. 14 (MJD 53596) at orbital phase 0.$^p$89. It nearly coincides with the peak of a minor outburst, with $\Delta U \approx 1.5$ mag, and $\Delta B \approx 1$ mag (Skopal et al. 2007).    Continuum points were chosen, in part, based on this spectrum in comparison with the $HST/STIS$ spectrum.  It has also been essential for understanding the profiles of the neutral and ionized helium  and Balmer lines.\footnote{We will present more details on these, and other, pre-outburst observations in a paper now in preparation.}  Classical photoelectric $UBVR_{\rm C}$ measurements in the standard Johnson-Cousins system were carried out by single-channel
photometers mounted in the Cassegrain foci of 0.6-m reflectors at
the Skalnat\'{e} Pleso and Star\'{a} Lesn\'{a} (pavilion G2)
observatories (see Skopal et al. 2004 for details).   
The star SAO\,16952 ($V$ = 9.88, $B-V$ = 0.56, $U-B$ = --0.04,
$V-R_{\rm C}$ = 0.32) and
SAO\,16935 ($V$ = 9.46, $B-V$ = 1.50, $U-B$ = 1.89)
were used as the comparison and check, respectively.
More recently $UBVR_{\rm C}I_{\rm C}$ CCD photometry was obtained
with the 0.5-m telescope at the Star\'{a} Lesn\'{a} Observatory
(pavilion G1). The {\sf SBIG ST10 MXE} CCD camera with the chip
2184$\times$1472 pixels was mounted at the Newtonian focus.
The size of the pixel is 6.8\,$\mu$m and the scale 0.56$\arcsec$/pixel,
corresponding to the field of view of a CCD frame about
of 24$\times$16 arcmin. Other details of the CCD photometric
reduction were described by Parimucha \& Va\v{n}ko (2005).
The same comparison stars were used as in Skopal et al. (2007).

 \section{Analysis}

The majority of our data set, listed in tables 1-2, consists of low and moderate resolution optical spectra, which include emission lines of the H Balmer series, He I/II, and O VI Raman features.    Cross calibration of the data sets is possible since we have overlap during late 2006 between the Loiano and the higher resolution Ond\v{r}ejov spectra, all of which were during either the decline or inter-outburst phases, and of all three sites in early 2007.  These permit a test of the effect of reduced spectral  resolution on the measurements of the narrow lines, particularly those of He I.  We find that the Loiano data systematically overestimate the equivalent width by about  20\% and, in creating the full time sequence, we include this correction in the plots.    There is essentially no effect for the broader lines.  The majority of the Loiano H$\alpha$ profiles are  saturated, but several obtained near optical maximum and during the first stages of the decline, taken with 10 sec exposure times, yielded reliable measurements and are included in Table 1b. Although Ond\v{r}ejov spectra missed the peak of the 2006 event, they  provide sufficient coverage of the subsequent evolution.   The equivalent width uncertainty was never above 10\%, being largest for the O VI 7080\AA\ feature and 5-10\% for all He I measurements.

The previously mentioned archival $HST/STIS$ medium resolution observation was absolutely calibrated. The fluxes and equivalent widths were measured for comparison with our data (units of : F(H$\alpha)$ = $7.51\times10^{-11}$ erg s$^{-1}$cm$^{-2}$  (EW = 111.6 \AA); F(He I 6678)=$1.27\times10^{-12}$ (EW=-2.08\AA), F(Raman 6825)= $4.41\times10^{-12}$ (EW=7.29\AA); F(He I 7065)=$1.64\times10^{-12}$ (EW=2.45\AA); F(Raman 7080)=$2.14\times10^{-12}$ (EW=3.44\AA).   The Raman 7080\AA/6825\AA\ line ratio was 0.47, higher than our outburst values but consistent with other observations taken during quiescence.  The 6825\AA\ line profile was similar to most quiescent spectra, with an intermediate equivalent width (see below).  

{\small 
\begin{center}
{\bf Table 1a. Journal of Observations\\
Loiano: He I/II and O VI Raman emission line equivalent widths} (\AA)\\
\begin{tabular}{cccccc}
\hline
JD & He II & He I  & O VI & O VI & He I  \\
2450000+  & 4686      &   6678  &    6825  & 7080 & 7065 \\
\hline
   4159.6 &     20.9   & 3.14 &        6.5 &      --- & --- \\
       4172.7 &    23.4 &    3.38 &        7.9 &       --- & --- \\
       4182.5 &     24.9 &   2.93 &        7.9 &       --- & --- \\
       4202.6 &    24.9 &    3.36 &        8.3 &       --- & --- \\
       4208.6 &     22.4 &   3.14 &        7.6 &       --- & --- \\
       4215.5 &    24.9 &    3.32 &        8.7 &      --- & --- \\
       4235.5 &     28.2 &   3.72 &        10.4 &       --- & --- \\
       4251.5 &    24.9 &    3.81 &        10.4 &       --- & --- \\
       4279.4 &   28.6 &     3.93 &        12.0 &       --- & --- \\
       4293.4 &     28.8 &   4.36 &        11.9 &      --- & --- \\
       4307.3 &    29.8 &    3.93 &        12.0 &      --- & --- \\
       4329.3 &     27.3 &  3.75 &        9.7  &       --- & --- \\
       4357.3 &   24.0 &     3.46 &        7.8 &      --- & --- \\
    \hline
       4008.3 &  --- &      4.20 &       $\le$0.2 &      --- & --- \\
       4033.2 &   --- &     3.81 &        2.4&     --- & --- \\
       4039.3 &    --- &    3.38 &        2.7 &       --- & --- \\
       4049.2 &    --- &    3.54 &        4.2 &       ---- & ---  \\
\hline
       4008.3 &   --- &     4.18 &       $\le$0.6 &      $\le$ 0.2 &       4.18 \\
       4033.2 &   --- &     3.50 &        2.0 &       0.7 &       3.69 \\
       4039.3 &  ---&       3.15 &        2.8 &       0.8 &       3.25 \\
       4049.2 &   --- &     3.34 &        4.3 &        1.6 &       3.49 \\
       4159.6 &     --- &   2.93 &        6.3 &        3.3 &       3.29 \\
       4172.7 &    --- &    3.22 &        7.8 &        3.5 &       3.47 \\
       4182.5 &    --- &    3.23 &        8.5 &        4.2 &       3.45 \\
       4202.6 &    --- &    3.53 &        8.1 &        3.9 &       3.36 \\
       4208.6 &    --- &    3.13 &        7.86 &        3.6 &       3.53 \\
       4215.5 &    --- &    3.16 &        8.8 &        4.0 &       3.48 \\
       4235.5 &     --- &   3.78 &        10.4 &        4.5 &       3.89 \\
       4251.5 &    --- &     3.81 &        10.9 &        4.5 &       4.06 \\
       4279.4 &    --- &    4.05 &        12.1 &        4.7  &       4.57 \\
       4293.4 &    --- &    4.15 &        12.0 &        4.7 &       4.53 \\
       4307.4 &   --- &     4.05 &        13.0 &        4.5 &       4.48 \\
       4329.3 &    --- &    3.81 &        10.4 &        4.0 &       4.08 \\
       4357.3 &   --- &     3.51 &        8.2 &        3.7 &       3.28 \\
\hline
\end{tabular}
\end{center}

\newpage

\begin{center}
{\bf Table 1b. Journal of Observations\\
 Loiano: Balmer emission line equivalent widths} (\AA)\\
\begin{tabular}{ccccc}
\hline
JD & H$\alpha$ & H$\beta$ & H$\gamma$ & H$\delta$\\
\hline
4008.3 & 102.8 & --- & --- & --- \\
4033.2 & 102.4 & --- & --- & --- \\
4039.3 & 113.7 & --- & --- & --- \\
4049.2 & 111.5 & --- & --- & --- \\
   4159.6 &   ---   &   21.69 &        10.82 &        5.70\\
       4172.7 &  ---   &   26.00 &        14.57 &        7.80\\
       4182.5 &   ---    &  27.79 &        16.21 &        10.01\\
       4202.6 &   ---   &  27.79 &        16.71 &        10.43\\
       4208.6 &   ---    & 26.88 &        14.93 &        8.03\\
       4215.5 &  ---    &  29.25 &        17.32 &        8.55\\
       4235.5 &   ---   &  35.61 &        20.42 &        11.82\\
       4251.5 &   ---   &  38.14 &        23.47 &        13.97\\
       4279.4 &  ---    &  44.11 &        27.03 &        16.04\\
       4293.4 &   ---   &  44.24 &        27.65 &        14.75\\
       4307.3 &  ---    &  46.12 &        28.70 &        16.48\\
       4329.3 &   ---   &  38.16 &        23.51 &        14.08\\
       4357.3 &  ---    &  29.88 &        16.78 &        8.98\\
  \hline
   4172.7 &    ---   &  25.93 &        13.42 &        7.23\\
       4182.5 &   ---  &   28.05 &        16.34 &        8.04\\
       4208.6 &  ---    &  26.47 &        14.90 &        8.71\\
       4215.5 &   ---   &  28.87 &        16.21 &        8.23\\
       4235.5 &  ---    &  34.66 &        21.09 &        10.84\\
       4251.5 &   ---   &  38.13 &        22.91 &        12.48\\
       4279.4 & ---    &  44.96 &        25.41 &        14.11\\
       4293.4 &   ---   &  43.84 &        25.21 &        13.79\\
       4307.4 &   ---   &  43.19 &        27.27 &        13.69\\
       4329.3 & ---    &  37.83 &        22.24 &        12.15\\
       4357.3 &   ---   &  28.77 &        16.04 &        8.13\\
\hline
 4293.4 &   --- &     43.62 &        27.48 &    --- \\
 \hline
 \end{tabular}
 \end{center}

\begin{center}
{\bf Table 2. Journal of Observations \\
Ond\v{r}ejov: Emission line equivalent widths} (\AA)\\
\begin{tabular}{cccc}
\hline
JD & H$\alpha$ & He I & O VI  \\
  2450000+  &   Wide    & 6678 & 6825 \\
\hline
 3145 &       76.1 &       1.26 &       --- \\
      3153 &       72.6 &       1.39 &       5.6 \\
      3454 &       77.7 &       1.06 &       --- \\
      4203 &       100.4 &       2.62 &       7.6 \\
      4217 &       106.4 &       2.51 &       8.1 \\
      4237 &       115.5 &       2.98 &       10.1 \\
      4266 &       136.3 &       3.34 &       11.0 \\
      4364 &       109.7 &       2.57 &       7.2 \\
       4557 &       124.1&      0.88 &       4.9 \\
      4595 &       91.4 &      0.96 &       --- \\
      4603 &       90.9 &      0.92 &       --- \\
      4614 &       83.6 &       1.11 &       --- \\
      4623 &       73.1 &      0.97 &       --- \\
      4627 &       78.0 &       1.02 &       3.9 \\
      4639 &       72.1 &      0.90 &       4.3 \\
      4645 &       76.7 &      1.02 &       4.0 \\
      4647 &       66.8 &       1.04 &       4.1 \\ 
      4648 &       75.9 &       1.00 &       --- \\
      4676 & 61.9 & 1.01 &  3.5 \\
      4685 & 50.9 & 1.03 & 3.7 \\
      4703 & 45.6 & 0.81 & 3.5 \\
      4718 & 41.1 & 0.85 & --- \\
      4761 & 41.1 & 1.04 & --- \\
      4924 & 65.9 & 1.80 & 4.7 \\
      5051 & 47.6 & 0.96 & --- \\
\hline
\end{tabular}
\end{center}
}

\subsection{O VI Raman features}

As mentioned in the introduction, AG Dra has one of the strongest O VI Raman pairs of any known symbiotic system.  The variations of these lines are discussed at length in, e.g. L04, G08.  Although the disappearance of the features in symbiotics has been noted sporadically in the literature (e.g. Tomov, Munari, \& Marrese 2000; Burmeister \& Leedj\"arev (2007) and references therein), to our knowledge this is the first time such an event has been observed in AG Dra or any other system during  an outburst that has been followed with sufficient coverage to provide details of the phenomenon.  The minimum line strength coincided with the optical peak of the outburst, although we have no spectra prior to optical (UBVR$_c$) maximum.   We show in Fig. 1 the first Loiano spectrum, on MJD 54007, and one taken near the end of the decline from the major outburst, MJD 54049.   The Raman 6825\AA\ feature was absent at the peak of the outburst although the He I 6678\AA\ and 7065\AA\  lines have about the same strength.  In Fig. 2 we show the variation of the equivalent width for the two Raman features compared with the $U$ and $V$ light curves.  In Fig. 3, we show the comparison of our equivalent width measurements with the published results of L04.  The 6825\AA\ feature varied in equivalent width by almost two orders of magnitude while the continuum varied by about a factor of 5; this is, therefore, not a masking effect of the continuum but a real, systematic variation of the line flux during outburst. 
    
    \begin{figure}
   \centering
   \includegraphics[width=9cm]{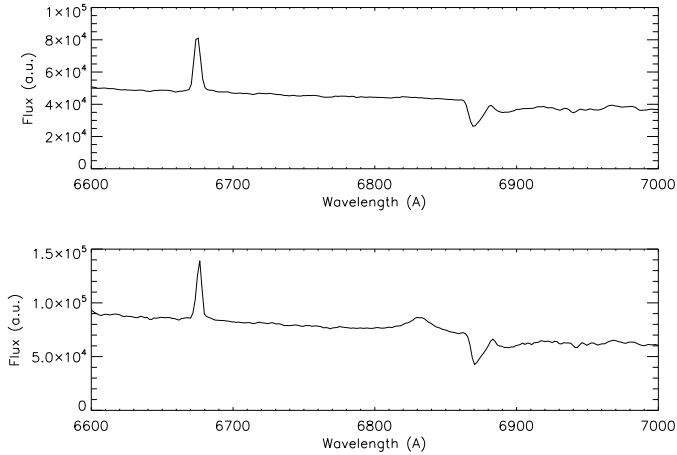}
   \caption{Comparison of Loiano 4\AA\ resolution spectra from 2006 Sep. 9 (top) and 2006 Nov. 9 (bottom).  The first taken at the peak of the outburst, shows the He I 6678\AA\ line about as strong as the second spectrum -- taken in the decline phase --  but with {\it no} detectable O VI Raman 6825\AA\ feature (see text for discussion).}
              \label{outburst-spectra}%
    \end{figure}

\begin{figure}
   \centering
   \includegraphics[width=8.5cm]{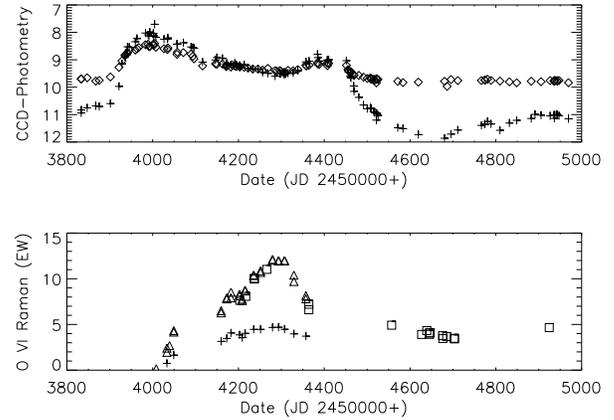}
   \caption{(top) U (plus) and V (diamond) CCD photometry for AG Dra throughout the observing interval; (bottom) Equivalent width (\AA) variations of the O VI Raman features: $\lambda$6825 (triangle, Loiano; square, 
   Ond\v{r}ejov) and $\lambda$7080 (plus, Loiano). }
              \label{raman-aavso}%
    \end{figure}
    The variation of the Raman 6825/7080 ratio, based on the equivalent widths, is shown in Fig. 4.    The change is statistically significant {\bf and systematic}, the first maximum corresponding to the maximum source brightness in our observing interval.    Since the last $FUSE$ spectrum from 2007 Mar. 15  nearly coincided with the absolutely calibrated Loiano spectrum of 2007 Feb. 27  we can make some further quantitative statements.   The flux ratio of the O VI $\lambda\lambda$1031, 1037 lines in 2007 was approximately the same as the optical Raman features, $\approx$0.45, which is considerably lower than the peak value that was also characteristic of the earlier $FUSE$ data.    Note that although the Raman line varied monotonically with visual magnitude during the 2006 outburst, this trend has not continued, especially during the weaker outburst in 2007 and only now is the emission returning toward its historical quiescent level.  Our most recent Ond\v{r}ejov data, from spring 2009, show that the $\lambda$6825\AA\ feature has remained below the minimum equivalent width reported in the time series in L04 while maintaining essentially the same profile as Fig. 5.   One additional measurement is available for the ratio for MJD 54703 from the Ond\v{r}ejov spectra,  EW(O VI Raman 7080)= 0.94\AA.  This corresponds to a 6825/7080 ratio of $\approx$3.7, the highest value we observed.  \\   

\begin{figure}
   \centering
   \includegraphics[width=9cm]{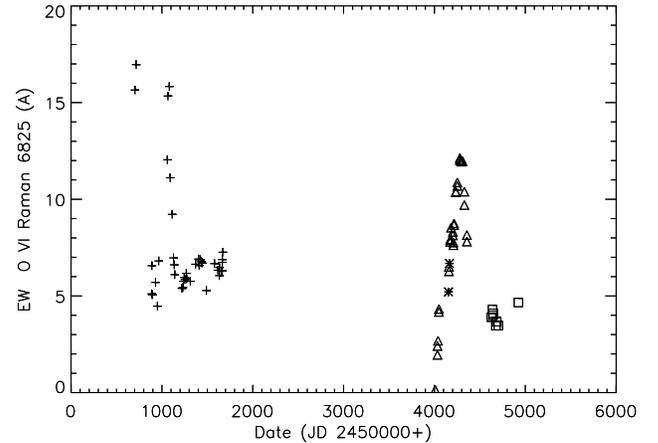}
   \caption{Comparison of the full data set of blue O VI Raman  $\lambda$6825 \AA\ feature equivalent width variations from the present outburst (triangle) and those reported in L04 (plus).  The H$\alpha$ interval was narrower for this comparison.  }
              \label{raman-long-term}%
    \end{figure}

    \begin{figure}
   \centering
   \includegraphics[width=9cm]{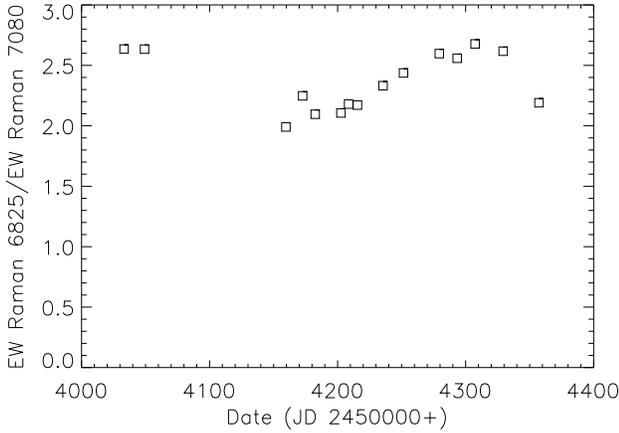}
   \caption{Variations of the O VI Raman $\lambda$6825$\lambda$7080 features from the Loiano data set.  The minimum  value is nearly the same as the $HST/STIS$ spectrum obtained during quiescence  (see text).}
              \label{raman-ratio}%
    \end{figure}
        
An interesting point that emerges from the $HST/STIS$ observation is the separation of the peak of the emission and the ``notch''  in the two features.  The velocity difference, $|\Delta v_{rad}|=220\pm7$ km s$^{-1}$, was quite similar despite the  different profiles.   This is compatible with from the Ond\v{r}ejov spectrum obtained during our interval when for the 6825\AA\ feature the separation was $\approx$200 km s$^{-1}$.  We discuss possible origins of this structure in the last section.    Figure 6 compares the line profiles for the O VI resonance lines from the last $FUSE$ spectrum with the Ond\v{r}ejov spectrum of the Raman features (MJD 54703).  The velocities have been corrected for the Raman effect wavelength shift (a factor of 6.6) in the optical spectra.   
 \begin{figure}
   \centering
   \includegraphics[width=9cm]{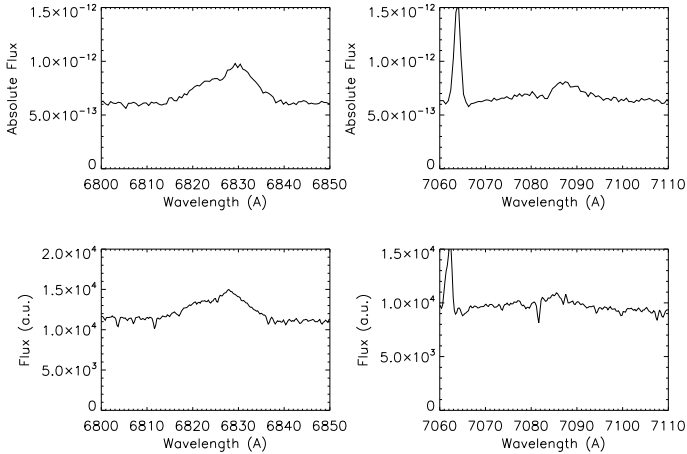}
   \caption{Comparison of O VI Raman line profiles for $HST/STIS$ spectrum (MJD 52755)  (top) and Ond\v{r}jov (MJD 54704)  (bottom).  The two spectra were obtained during similar quiescent epochs of AG Dra.  The strong emission line blueward of O VI 7080\AA\ is He I 7065\AA.}
              \label{FigGam}%
    \end{figure}
\subsubsection{Raman conversion efficiency}

We attempted to use the last $FUSE$ data to determine the Raman conversion efficiency that, for AG Dra, has been studied based on data taken in the 1990s.  For the Raman features, the measured fluxes from Loiano were F(6825)=4.45$\times 10^{-12}$ erg s$^{-1}$cm$^{-2}$ and F(7080)=1.94$\times 10^{-12}$.  It should be noted that these are almost identical to the values from the $HST/STIS$  spectrum, see above.  For the $FUSE$ spectrum, the {\it measured} fluxes were F(1031)=1.7$\times$10$^{-12}$ erg s$^{-1}$cm$^{-2}$ and F(1037)=0.7$\times 10^{-12}$.   Thus, the doublet ratio was similar for the O VI lines, 0.44 for the Raman features and 0.42 for the FUV doublet.   The $FUSE$ fluxes are, however, not directly usable for deriving the Raman conversion efficiency, $Y=N(6825)/N(1031)$,  based on the comparison in both line and continuum fluxes and the Science Data Assessment form for observation D0090106.  If, however, the continuum fluxes scale then it is possible to at least place a limit on $Y$.  Using the Hutchings \& Gaisson (2001) and Sofia et al. (2005) FUV extinction curves for the Galaxy, and the mean visual extinction E(B-V)=0.11 from  Schegel, Finkenbeiner, \& Davis (1998) maps, the extinction correction for AG Dra at the FUV O VI doublet is 6.9.   The spectrum was taken around quiescence following the major outburst.  If the actual continuum level was approximately the same, we find an additional correction factor of 6$\pm$1 for the continuum level and $Y \approx 0.5$ and $0.4$ for the two components, respectively.  If, instead, we adopt E(B-V)=0.08 (see Young et al. (2005) and Birriel, Espey, \& Schulte-Ladbeck (2000)) we find $Y \approx 0.4$ and $0.3$ for the two components applying the differential H$_2$ absorption correction factors from Birriel et al. (respectively 0.97 and 0.64 for the two components of the FUV doublet).

    \begin{figure}
   \centering
   \includegraphics[width=9cm]{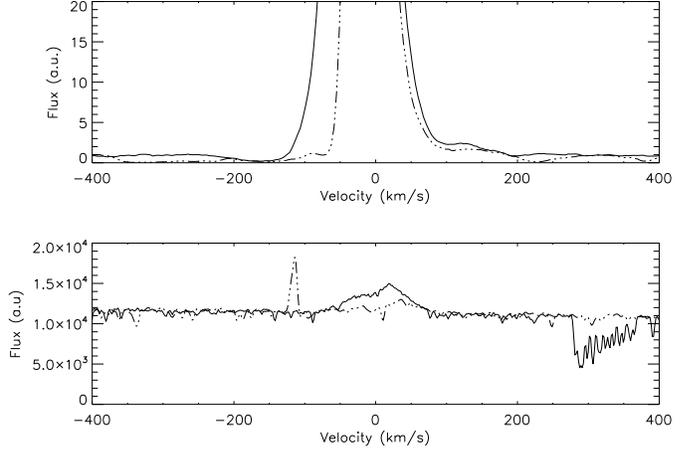}
   \caption{Comparison of  the $FUSE$ O VI doublet spectra (D0090106, 2007 Mar. 15) (top panel: 1031.9\AA\ solid,   1037.6\AA\ dot-dash) with the Ond\v{r}ejov spectrum shown in Fig. 5 (bottom panel: 6825\AA\ solid, 7080\AA\ dot-dash).  Fluxes are normalized, and the radial velocities for the optical lines have been scaled by the Raman frequency shift  (see text for discussion).  The narrow emission feature in the bottom panel is He I 7065\AA.}
              \label{ond-FUSE}%
    \end{figure}
 
This tentative efficiency is similar to that obtained by Schmid et al. (1999), $Y$=0.5 for $\lambda$6825 and between 0.3 and 0.5, with an uncertainty of order 50\%,  from  the two observations of $\lambda$7080 using $ORFEUS$ data obtained in 1993 and 1996; neither was obtained during an outburst.  In contrast,  this value  is  larger than found by Birriel et al.  (2000), who obtained $Y \approx 0.14$ for the two Raman features using contemporaneous $HUT$ and groundbased spectra from 1995 Mar 16 when the source was, according to AAVSO data,  declining  after a sequence of  outbursts.  The ratio of the O VI FUV doublet lines during the $HUT$ observations, 3.23, is quite different from ours and about the minimum value obtained during the $FUSE$ era.  We do not have coincident FUV data for the peak of the outburst when the O VI Raman feature was undetectable.   

\subsubsection{O VI Raman line profiles}

The O VI $\lambda$6825 line profile appears to consist of two {\it components} that vary similarly when the Raman lines are strong.  The red component is significantly enhanced relative to the blue, with the ratio of their peaks reaching up to $\approx 2$.  As proposed by Schmid et al. (1999), we interpret this as two separate contributors, one from the wind (and possibly  accretion disk) of the WD, the other from the ionized  region surrounding it within the K star wind.  The displacement in the centroid is consistent with the ionized region having a different velocity than the WD wind, the result of a flow through the ionization zone.  The centroid of the line is shifted by approximately +300 km s$^{-1}$, again consistent with scattering from a wind with a velocity of order 50 km s$^{-1}$ relative to the WD.  Although the intrinsic Raman profiles are not Gaussian, to parameterize the shape variations we decomposed the Raman 6825\AA\ using two Gaussians.  The EW variations for the two components are shown in Fig. 7.  The choice of profile was a convenience, in the absence of the FUV doublet we chose this as a way of displaying the relative contributions of the two components.  The redward component is always stronger.  The minimum value of the ratio in 2004 is about the same as the post-outburst value, 0.27, while the peak of the ratio occurs  at the peak of the minor outburst and at Balmer line minimum.  We emphasize, however, that this is a {\it phenomenological} characterization of the profile variations, it is also plausible that there is only a single component and the changes reflect, instead, variations in the O VI resonance line profile at 1031\AA\ due to varying P Cyg absorption on the blueward side of the profile.  We do not have sufficient coverage of the 7080\AA\ Raman feature  to compare the variations but, as shown above,  the profiles are not the same.  The stronger  component, at longer wavelengths, is narrower in velocity and more variable than the broad component.  

\begin{figure}
   \centering
      \includegraphics[width=9cm]{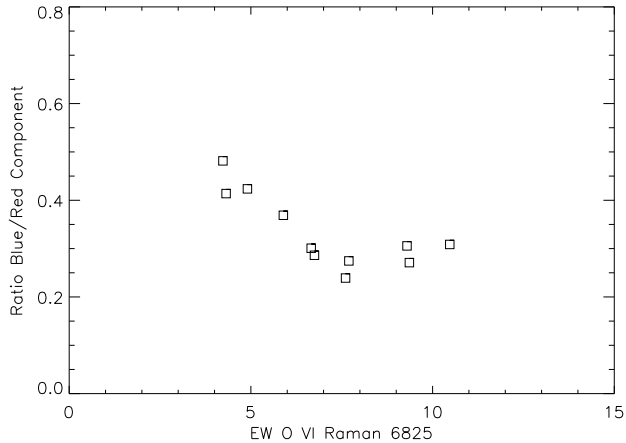}
   \caption{Ratio of the equivalent widths of the two O VI Raman 6825\AA\ components obtained by Gaussian decomposition as a function of the line strength (in \AA).  The blue component is both weaker and consistently broader, the broadest profile corresponding to the strongest integrated flux and optical minimum.}
              \label{raman-component-ratio}%
    \end{figure}

For the interval 1997 to 2003, L04 report a correlation of H$\beta$ emission equivalent width with orbital period.   For comparison, using a different measurement, we show, in Fig. 8, the variation of the O VI  FUV doublet and optical Raman line ratios  phased on the radial velocity ephemeris (Fekel et al. 2000).  The FUV spectra were obtained mainly in quiescence, while our spectra were taken during outbursts so any search for orbital modulation of the line strengths is hindered by the outburst, which lasted about 0.$^p$2.  This dominates the relatively brief  time interval of these equivalent width measurements  and we regard any variations  as being due to the outbursts.  A dedicated monitoring of this system is needed to assure that a dense enough coverage is available during any quiescent phase to disentangle the effects.   If the orbital modulation of the line ratio is real, being a single wave with minimum at K star elongation may indicate a slight eccentricity in the system or an asymmetry in the WD environment.

      \begin{figure}
   \centering
   \includegraphics[width=9cm]{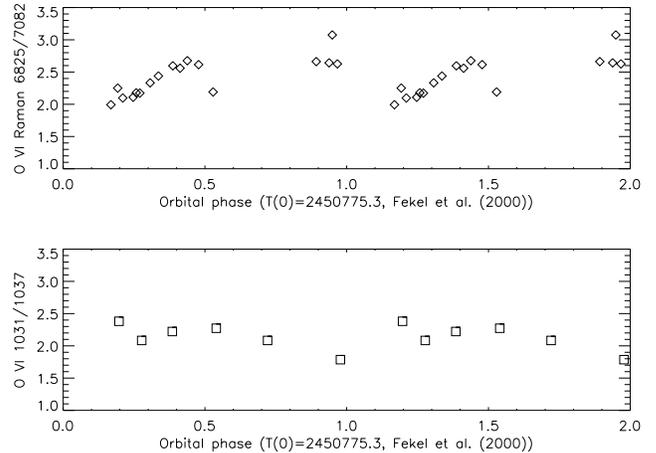}
   \caption{Flux ratio of the FUV O VI doublet components (bottom) and the optical Raman features (top) as a function of orbital phase based on the Fekel et al,. (2000) radial velocity ephemeris; phase 0.$^p$0 being elongation of the K star.}
              \label{raman-ratio-phased}%
    \end{figure}

\subsection{Helium lines}

Figure 9 shows the variation of the helium lines through the outburst.  The He I 7065\AA\ line showed similar variations but our coverage is not as complete.  To highlight the qualitative change in the AG Dra variations, we also show the comparison between the Raman 6825\AA\ feature and He I 6678\AA\ (lower right panel).  This highlights the difference between the outburst and post-outburst variations, {\bf already visible from the time series}.  The difference is real, the same relative behavior is seen in the H$\alpha$ variations.  This does not appear to be orbital modulation (that would not seriously affect optical lines).  Instead, it appears that the excitation significantly changed while the He II 4686\AA\ line did not show any departure from its earlier behavior.  High resolution TNG  data, of which we show a sample in Fig. 12 (to be discussed more extensively in a separate paper, in preparation), show absorption on all He I line profiles at about -50 km s$^{-1}$ (relative to the systematic velocity of -144 km s$^{-1}$) at the same velocity as the Balmer line absorptions.  But at most, the variation in the absorption component accounts for 20-30\% of the equivalent width variations.  No such feature is seen on the He II 4686\AA\ profile whose peak corresponds to the peak of the Balmer lines.  The He II, at high resolution, is asymmetric with an extended red wing, perhaps indicating a weak absorption trough on the approaching side.

       \begin{figure}
   \centering   
   \includegraphics[width=6.5cm,angle=90]{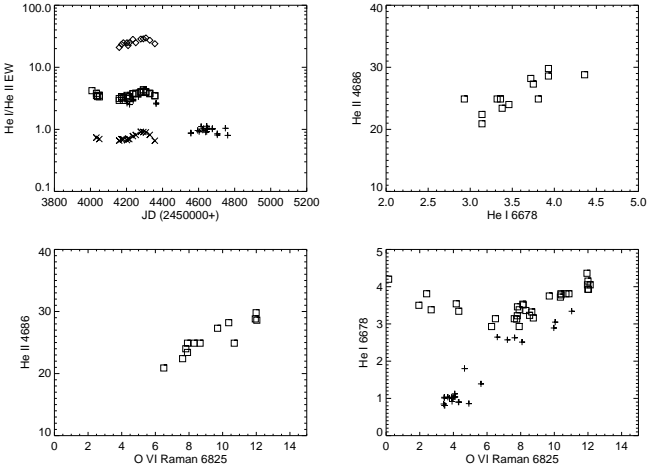}
   \caption{Equivalent width variations of the He I and He II lines.  Upper left: He II 4686 (diamond, Loiano only), He I 6678 (square, Loiano; plus: Ond\v{r}ejov) and He II 7065 (cross, scaled by 1/5, Loiano only); upper right: He I 6678\AA\ vs. He II 4686\AA\ (Loiano); lower left: O VI 6825\AA\ vs. He II 4686\AA\ (Loiano only); lower right: O VI 6825\AA\  vs. He I 6678\AA\ (square: Loiano, plus, Ond\v{r}ejov).  See text for discussion.}
                 \label{6678-4686}%
    \end{figure}
    
The strongest He I lines ($\lambda\lambda$5875, 6678, and 7065\AA) all show a peak coinciding with the He II and Raman features, at the minimum of the visual flux.  As reported by L04 and G99, we find $<$EW(He II 4686)/EW(H$\beta$)$>$=0.80$\pm$0.17; at no epoch did this exceed 1.0, but we emphasize that these spectra did not coincide with the peak of the major event.  Notably, none of our spectra displayed a He II 4686\AA\ line as strong as the most extreme reported by L04.  The line strength during the declining phase of the outburst was strongly correlated with that of the Raman 6825\AA\ feature but at line minimum light appears to vary relatively less.  There is, in addition, a single late measurement, MJD 54703, of He I 7065\AA\ from Ond\v{r}ejov, EW(He I 7065) = 0.89\AA, which is the weakest recorded during this observing interval.  Birriel  (2004) reports the detection of Raman scattered He II in the spectra of HM Sge and V1016 Cyg.  We find no evidence for  this feature in any AG Dra spectra, neither the lower resolution Loiano data taken during the outburst nor the high resolution TNG spectra taken outside of outburst.
    
The He I lines vary in phase with the He II $\lambda$4686 line during the decay phase of the outburst; we have no simultaneous  spectra at peak but there is an additional, weak line is present at 6680.2\AA\, which coincides with the NIST listing for O IV 6682\AA\ (at the systemic velocity) but it is more likely He II 6682\AA.  This line varies but we do not report those results since the S/N ratio is not sufficiently high in most of the spectra to provide accurate measurements.  Its properties are the same as the helium lines, and we thus identify this as a recombination line from the same ionized region.  It appears to be constant in the Ond\v{r}ejov spectra, where it is barely detectable; because of the low resolution and SNR of the $HST/STIS$ spectrum it is not visible.   This line is visible in the velocity plot for He I 6678\AA\ (Fig. 12).
  
\subsection{ Balmer lines}

All quoted equivalent widths were measured between $\pm$2000 km s$^{-1}$ for consistency with previous reports of extended wings and based, in part, on continuum points selected with the TNG and $HST/STIS$ spectra.   The extended wings (from 1000 to 2000 km s$^{-1}$) contribute $\approx$25\% of the total flux.  Fig. 10 shows the variations of the equivalent widths; our coverage of the major outburst only includes H$\alpha$.
       \begin{figure}
   \centering   
   \includegraphics[width=9cm]{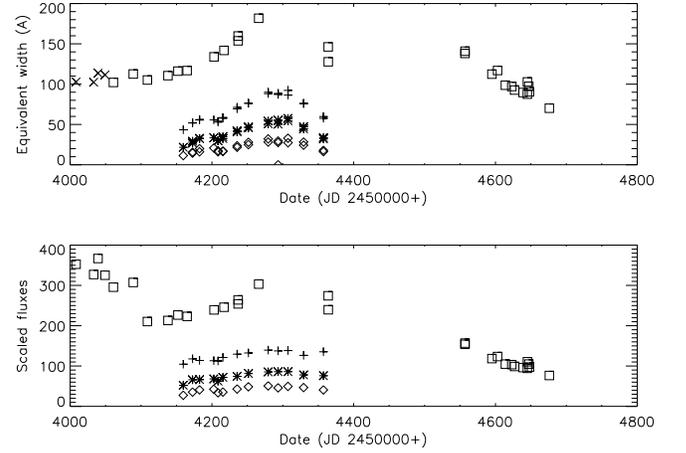}
   \caption{Variation of the Balmer lines from the beginning of the 2006 outburst of AG Dra; H$\alpha$ square: Ond\v{r}ejov, cross: Loiano.  All other Balmer line measurements are only from Loiano spectra --  H$\beta$, plus; H$\gamma$, asterisk; H$\delta$, diamond.  Top panel: equivalent widths (\AA), bottom panel: scaled fluxes (see sec. 3.5, below).}
              \label{Balmer-EW}%
    \end{figure}
Our coverage of the post-outburst period shows that H$\alpha$ remained below its level during the outburst and decline phase.  The minimum value we find is consistent with that reported in L04 (their Fig. 4 shows the variations phased according to the photometric ephemeris).   Their equivalent widths for both H$\alpha$ and H$\beta$ are greater than any we observed during the recovery from the major outburst .  The effects on the interpretation of the He  line variations due to the continuum variations are, however, important and we will return to this point presently.

\subsection{High resolution Balmer and He I line profiles}

 \begin{figure}
   \centering   
   \includegraphics[width=9.2cm]{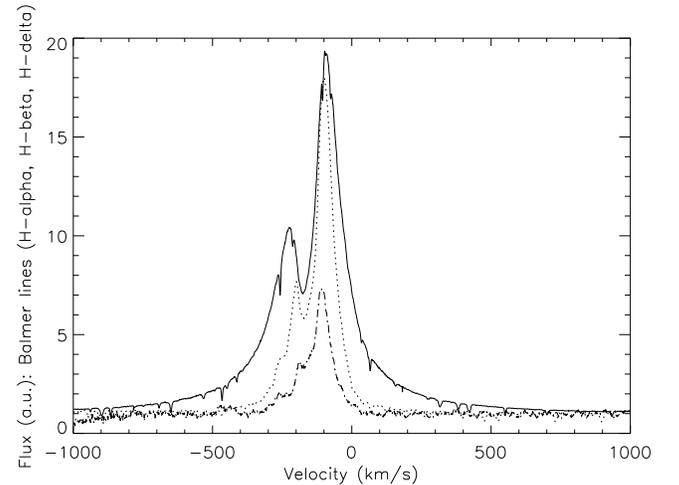}
   \caption{Sample Balmer line profiles, TNG spectra,  2005 Aug. 14 (MJD 3596).  Solid: H$\alpha$, dotted: H$\beta$, dashed: H$\delta$}
              \label{Balmer-profiles}%
    \end{figure}

The TNG  spectrum, with a resolution of  $\approx$0.1\AA, is capable of resolving the structure on the neutral helium emission lines; none of the Ond\v{r}ejov or Loiano spectra have sufficient resolution to detect this, nor were they resolved in the medium dispersion $HST$/STIS spectra.   The He I lines are all well resolved and, more significantly, it display  complex profiles with an extended red wing and FWZI = 150 km s$^{-1}$;  The absorption  feature is at virtually the same radial velocity for the three lines,  -158$\pm$3 km s$^{-1}$, while the peak velocity is -125 km s$^{-1}$ for He I 5875, -143 km s$^{-1}$for He I 6678, and -130 km s$^{-1}$ for He I 7065.   The H$\alpha$ line shows strong absorption at -177 km $^{-1}$,  but with the peak velocity displaced to -92  km s$^{-1}$, about 50 km s$^{-1}$ to the red of the systemic velocity.  Absorption is present on all He I lines regardless of multiplicity; as shown in Fig. 12,  the absorption decreases in order of   He I $\lambda\lambda$5875,  7065, and 6678.   All  He I lines are extremely narrow compared  to the Balmer profiles, FWHM(He I) $\approx$ 80 km s$^{-1}$.  In contrast, the He II 4686\AA\ profile is single peaked at the system velocity with a FWHM  = 70 km s$^{-1}$ in the TNG and the $HST/STIS$, spectra,  although taken at activity states of the AG Dra.   These are consistent with the lines being formed near the WD but with a puzzle regarding the cause of the absorption.  It cannot  be from the wind of the red giant and must, instead, arise within the embedded ionized region around the WD   In the 
Ond\v{r}ejov spectra, the H$\alpha$ profiles display variable absorption (e.g. Smith \& Bopp 1981, Ivison et al. 1994, and Belczynski et al. 2000).   We reserve further analysis of these variations for the next paper.
     \begin{figure}
   \centering   
   \includegraphics[width=9.2cm]{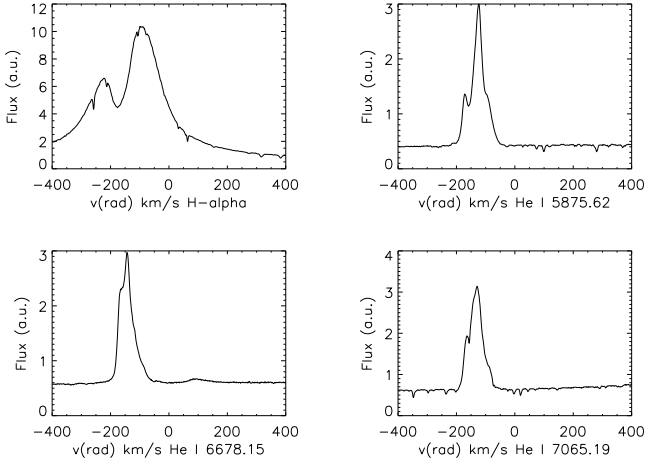}
   \caption{Comparative line profiles for H$\alpha$ and the red He I lines  in the TNG spectra,  2005 Aug. 14 (MJD 3596).  The weak feature to the red of He I 6678, at $v_{\rm rad} \approx +110$ km s$^{-1}$, is identified as He II 6682.  See text for discussion.}
              \label{Ha-He-vel}
    \end{figure}
     \subsection{Line flux variations based on broadband photometry}
          
With the exception of the major outburst, line variations can be appropriately characterized by equivalent widths alone.  However, for the outburst interval, from MJD 4000 to 4400, the continuum variations alter the interpretation of the line strengths for the Balmer and helium lines.  We have adopted a scaling  based on calibrated  spectra taken outside of outburst along with the two in this sequence, from Loiano at the extremes of the major outburst (Sept. 2006 and Feb. 2007) and in the B, V, and R$_C$ photometry.  {\bf Assuming that the visible continuum and emission lines are independently varying, we correct the continuum level by dividing by the continuum flux (based on the photometry), normalized to the minimum V and R magnitudes (using V and  $\Delta R_c$ light curves) around 2007 Feb. 15, roughly coincident with the last $FUSE$ spectrum and a date for which we have a flux calibrated red spectrum.}    In Fig. 10 (bottom panel) we show the effect on the Balmer lines, in Fig. 13 for the Raman 6825\AA\ feature,  in Fig. 14 on the He I 6678\AA\ line, and in Fig. 15 for He II 4686\AA.   As discussed above, the Loiano spectrum from MJD 54159 was absolutely calibrated.  The Balmer series fluxes were, in units of $10^{-12}$erg s$^{-1}$cm$^{-2}$, F(H$\alpha$)$>$62 (saturated), F(H$\beta$)=10.4, F(H$\gamma$)=3.67, F(H$\delta$)=1.94.  For the helium lines, in the same units, F(He I 5578) =2.54, F(He I 6678)=1.95, F(He I 7065)=2.51, and F(He II 4686)=9.04.  All fluxes quoted are uncorrected for the (minimal) visual reddening.   
 \begin{figure}
   \centering   
   \includegraphics[width=9.2cm]{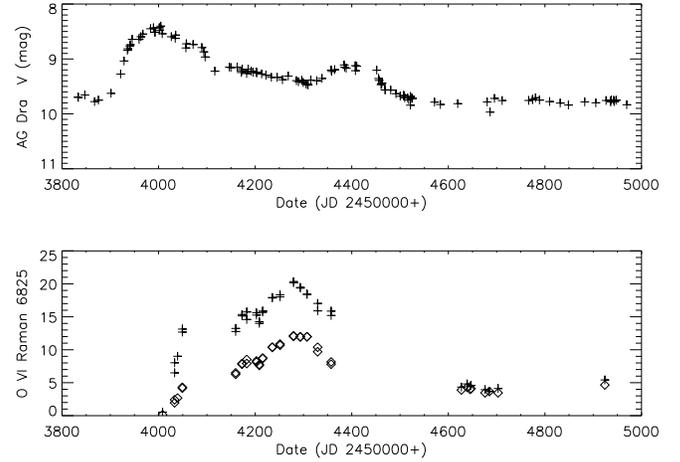}
   \caption{Top: CCD-V magnitudes for AG Dra during outburst and recovery (see text); Bottom: Variations of the O VI Raman 6825\AA\ feature with and without correction to normalized fluxes based on the optical light curve.  Diamond: combined 
   Loiano/Ond\v{r}ejov equivalent widths; plus: scaled fluxes generated based on the V photometry.}
   \label{corrections0}
    \end{figure}

 \begin{figure}
   \centering   
   \includegraphics[width=9.2cm]{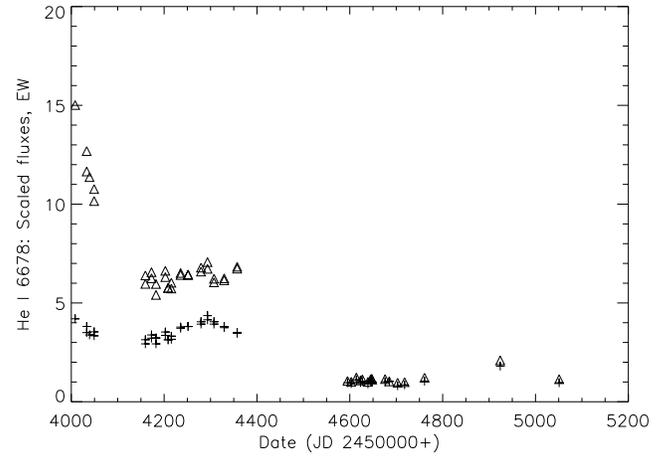}
   \caption{Comparative variations of the equivalent widths of the He I 6678 line with and without correction to normalized fluxes based on the optical light curve.  Triangles, squares: Loiano/Ond\v{r}ejov scaled fluxes; plus, equivalent widths uncorrected .}
   \label{corrections}
    \end{figure}

 \begin{figure}
   \centering   
   \includegraphics[width=9.2cm]{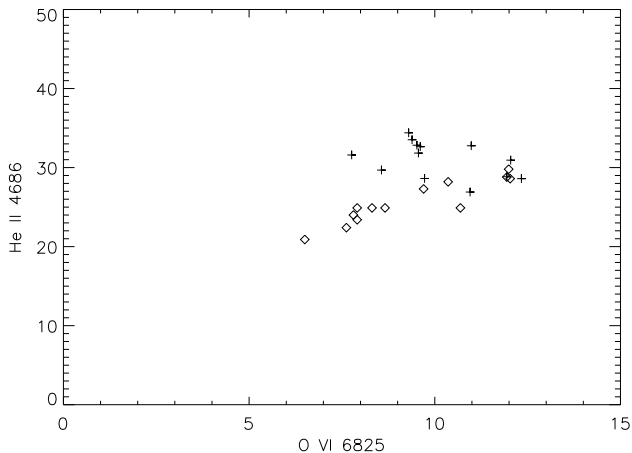}
   \caption{Comparative variations of the equivalent widths of the O VI Raman 6825\AA\ line with He II 4686\AA\ based on the B and V photometric variations.  Plus: equivalent widths; diamond: scaled fluxes.}
   \label{corrections1}
    \end{figure}

The main effects of the correction are to reduce the interval of the weak Raman emission phase and to change the variations of He II 4686\AA\ relative to the He I lines.  During the maximum of the major outburst, the {\it upper} limit for the EWs of both  Raman lines is more than an order of magnitude lower than the strongest emission (quiescence).  During the decline phase the ratio of the two Raman features remains approximately constant.  As reported by L04, the He II 4686\AA\ line is positively correlated with the Raman emission but once the correction is applied for the continuum variations this is not as tight a relation, a result of the correlation induced by the continuum variations.  The neutral helium lines vary in phase with He II 4686\AA\ at the strongest phases (Fig. 9).  The variations of the Balmer lines  are largely unaffected since our Loiano data for H$\beta$-H$\delta$ cover only the decline and quiescence stages of the outburst.
     
\section{Discussion}

Our results both confirm and extend previous studies of the line variations in AG Dra, especially L04 and G08,  but also present important novelties.  The most significant is the vanishing -- and recovery -- of the Raman lines during and following the peak of the major outburst.  Since there is no indication that the wind of the red giant has significantly changed in velocity or density, the hydrogen line variations are consistent with the same profile although the line flux decreases, it is more likely that this signals the disappearance of the emission component of the O VI resonance lines.  This observation provides further support for the Raman line as a proxy measure of the state of the highest ionization region and inner fast wind around the gainer.  

The minimum Raman line equivalent width during the decade before the 2006 outburst was about 6 \AA,  about the same as  our first Ond\v{r}ejov spectrum and the $HST/STIS$ measurement.  During the entire recent outburst, we observed  consistently  lower values. This can not be due to continuum variations alone, although some may be present at about the 20\% level based on the variations of the He I lines.    If each of the Raman features consists of two components, as suggested by the decomposition, then the line variations indicate that both the WD wind and ionized region disappeared during the peak of the major outburst.  Since we do not have observations prior to the photometric peak we cannot constrain the density from the time dependence alone.  The recovery of the emission line was more rapid than the rate of continuum decrease, seen from the comparison of the light curve with the Raman flux variations.
    
The suggestion that the soft XR and UV variations are due to expansion and falling effective temperature of the gainer in this system receives support from the behavior we observe (Greiner 1997, Gonzalez-Riestra et al. 1999).  The WD develops an optically thick, extended atmosphere with a large covering factor that suppresses the {\it resonance O VI lines} and the ionized region but maintains sufficient excitation to produce the emission at He I. The conversion efficiency for UV photons by Raman scattering has been obtained from previous observations of the system during quiescence and unless there is a drastic change in the structure of the wind in the accretion region we would expect it to remain roughly constant, so the disappearance of the optical Raman lines indicates an increase in the optical depth of the UV and the redistribution of the radiation to longer wavelengths. If the redistribution is essentially passive -- that is, if during the outburst the bolometric luminosity of the WD does not significantly change -- the maximum increase we would expect in the visible is about one magnitude. The brightest outbursts are substantially larger than this, using $U$ and $B$ as the proxies for the redistribution since the red giant contributes most of the flux at $V$ and longer wavelengths. Instead, if a thermonuclear runaway is initiated on the gainer, the increase in the radius of the pseudo-photosphere will produce a substantial change in the ultraviolet spectrum, including the appearance of neutral and low ionization state lines of the iron peak (the so-called ``iron curtain'' well known from classical novae and Luminous Blue Variables) and the disappearance of the highest ionization wind lines that should normally be visible from the gainer. This antiphase behavior, so well known from LBVs, is consistent with $IUE$ and $ROSAT$ observations reported by Greiner et al. (1997) and G08 during the inter-outburst period preceding the latest eruption.  To increase the optical depth sufficiently in the UV requires column densities of order $10^{24}$cm$^{-2}$ or higher and would be consistent with an expansion of the WD photosphere accompanied by an optically thick wind.

Lee and Kang (2007) propose an accretion disk as the source for Raman scattered photons  and model the inequality of the peaks as an optical depth effect by self-absorption, in the RG wind.   It is possible that this is a contributor to the spectra, hydrodynamical modeling of symbiotic wind does produce disks (e.g. Walder, Folini, \& Shore 2008), but it is far more direct to decompose the profile into one from the scattering of the WD wind and the other from the surrounding ionized nebula of the K star wind.   The intensity ratio of the presumed two peaks is slightly variable in the Ond\v{r}ejov spectra, depending on the line strength.  The Raman 6825\AA\ line profile is consistent with the interpretation by Schmid et al, (1999), based on the comparison of $ORFEUS$  and optical spectra, that the source photons arise from the ionized region around the hot component and are scattered in the slow wind of the K giant.  This would imply that the timescales for the variations should be of order the recombination time and that the outburst, extinguishing the FUV and soft XR, causes a collapse of the O$^{+5}$ region around the WD.  Additional changes on a shorter timescale, mainly in the broader component, could be due to the same mechanism that produces the outburst, that the cooling of the pseudo-photosphere and the increase in the fast wind optical depth emission from the Raman lines.  The 7080\AA\  feature has a different profile.  We suggest that in addition to the intrinsic WD wind absorption component, and that of the ionized region in the RG wind, there are additional absorbers -- both in the ionized and neutral K star wind -- that are altering the Raman profile.  Several multiplets of S I ground state transitions between 1030 and 1038\AA\ and the C II 1036, 1037\AA\ doublet may not be entirely interstellar in origin.  Instead, any additional absorption would not only alter the Raman line profile but also affect the line ratio.  If, for instance, during a major outburst the C II absorption line varies, this will affect the conversion efficiency and the resulting optical profiles.  Since, however, the region is heavily blanketed by interstellar H$_2$ it is not possible to obtain an absorption equivalent width and lacking contemporaneous pairs of FUV and optical spectra during the peak of the outburst, this must remain an hypothesis.

The timescale for an outburst, about four months, is approximately the recombination time for a region with an electron density of about $10^8$cm$^{-3}$ for the O$^{+5}$ zone, using the recombination coefficients from Nahar \& Prasad (2003) in the temperature range 15 - 100 kK and consistent with the previous electron density estimates in the literature, e.g. Young et al. (2005, 2006).  This would explain the absence of the [O III] lines in a region sufficiently ionized to otherwise show them.  The He II 6682\AA\ line, weakly present in the best exposed spectra, is slightly stronger relative to He I 6678\AA\ in data from 2008 June, possibly indicating ionization has occurred.  The data from the main outburst do not have sufficient resolution to show this line, and there are no contemporaneous UV spectra during a similar phase (although the one $FUSE$ spectrum, taken almost at quiescence, shows the O VI resonance lines at about their maximum strength).  This may explain the weakening of the red component of the Raman line.   The disappearance of both components of the Raman features also challenges shocked wind models for events during the outburst that must produce temperatures lower than those of the photoionization source.  The increase in the visual flux, 0.3-1.0$\mu$, and the contemporary decrease in the ultraviolet are compatible with  simple flux redistribution from an optically thick pseudo-photosphere mentioned above.  This would easily account for the disappearance of the high ionization features and the persistence of the He II and He I emission during optical maximum.  Further discussion of the long term history of all emission lines and modeling of this system is postponed to the next paper.

{\bf Note added in revision: A parallel study of the 2006-2008 outburst was published after we had submitted our paper (Munari, U. et al. 2009, PASP, 121, 1070) that reports photometric and low resolution spectroscopic observations of the event.  Their spectrum from near the peak of the outburst, 2006 Sep. 30 (MJD 54008) is simultaneous with ours  at the peak of the outburst and confirms the absence of the two O VI Raman features, identifying the event as a ``cool outburst'' in the sense of G08.  Their subsequent spectra confirm the recovery of the line.  They also report, based on an echelle spectrum (R=35000) taken on 2006 Oct. 6,  the presence  of P Cyg components on the He I lines with velocities  similar to those we describe here.  A further discussion of these results will be included in our next paper.}

\begin{acknowledgements}

Research at  Loiano is supported by INAF.   PK was supported by ESA PECS grant No 98058. GMW acknowledges support from NASA grant NNG06GJ29G. We thank the staff at Loiano for their kind help in obtaining many of the spectra used in this study as a service observing program and U. Munari, C. Rossi, and R. Viotti for use of unpublished spectra.     Some spectra at Ond\v{r}ejov were taken by M. Netolick\'{y},  B. Ku\v{c}erov\'{a}, V. Votruba and D. Kor\v{c}\'{a}kov\'{a}.    MW was supported by the Research Program MSM0021620860 of the Ministry of Education of the Czech Republic.    AS was supported by  a grant of the Slovak Academy of Sciences No. 2/7010/27.  We thank T. B. Ake for collaboration on the $FUSE$ proposal that began this observation, unfortunately not executed before the end of the mission, and for his advice on the $FUSE$ data, and  J. P. Aufdenberg, C. Rossi, and R. Viotti for valuable discussions.  The $HST/STIS$ and $FUSE$ spectra were obtained from the MAST archive of STScI and archival visual photometric data were provided by the AAVSO.   
 
\end{acknowledgements}

\end{document}